# NANOMODIFICATION OF ALUMINUM ALLOYS

A. Titlov[1], G. Krushenko[2,3], S. Reshetnikova[3], K. Borodianskiy[4]

[1]Odessa National Academy of Food Technologies, Odessa, Ukraine
titlov@cloud.onaft.edu.ua
[2] Institute of Computational Modeling SB RAS, Krasnoyarsk, Russia
[3] Siberian state University of Science and Technology
named after academician M. F. Reshetnev, Krasnoyarsk, Russia
[4] Ariel University, Ariel, Israel

**ABSTRACT.** This work is an overview of aluminum alloy crystallization mechanism modified by different refractory nanopowders.

R. Feynman, a well-known physicist and a Nobel Prize winner, is considered as a founder of nanotechnology. In 1959, in his famous lecture known as "*There's Plenty of Room at the Bottom*" [1], he outlined a great potential of design on a molecular and atomic scale, which can be achieved by obtaining materials and devices in atomic-molecular range. In this lecture, he also noted that a new type of those structures will have to be developed to control the properties of such tiny nano-structures.

In the early 80-s, a MIT engineer K.E. Drexler highly contributed to the development of nanotechnologies by proposing his own methods and physical principles of obtaining nanoscale complex structures [2].

Norio Taniguchi from the Tokyo University is considered to be the first to define the term of "nanotechnology" in 1974. When speaking at the conference of Japanese Society for Precision Engineering and Nanotechnology about an imminent transition to high-precision treatment [4], he defined it as a technology that ensures an extra high precision and ultra fine dimensions of about 1 nm. Prof. Taniguchi described nanotechnology as follows: "Nano-technology mainly consists of the processing of separation, consolidation, and deformation of materials by one atom or one molecule".

Many governments, especially in the developed countries, believe that nanotechnology can become one of the key factors of economic growth. Commercial use of nanotechnologies will be a key component of economic growth and a means of restructuring of national economies in strategic industries. According to Lux Research (USA) [5], which is a leader in analyzing the world nanotechnological markets and works in a close cooperation with governmental bodies and largest private corporations worldwide, an average price on nanotechnology-based products currently reaches 11% as compared to similar conventional materials and devices.

A long-term national program called "National Nanotechnology Initiative" (NNI) was adopted in the USA at the end of 2000. This program is targeted to financing research and development in nanoscience and nanotechnology, with 14.2 billion USD allocated to implement it during 2001 - 2011.

Over the last 11 years, the governments in 60 countries worldwide have invested over 67 billion USD in R&D in the field of nanotechnology [6]. A special attention in this regard is paid to refractory nanopowders (NP) (such as nitrides, carbonitrides, carbides, borides, oxides, etc.). Usually those nanopowders used to improve physical and mechanical properties of various metallic materials [7].

NPs are ultrafine-grain crystal or amorphous substances which are as small as 100nm (1nm = $1 \times 10^{-9}$ m). They have unique physical, chemical and mechanical properties, which are very different from those of large scale materials. This uniqueness is attributed to the fact that the number of atoms in the surface layer and in the bulk is commensurable. As the atoms in the surface layer of nanoparticles have their counterparts from one side only, their balance is impaired, and the resulting structure relaxation leads to the displacement of atomic spacing in a 2…3nm layer. Therefore the surface layers expand while internal layers compressed, as these are exposed to an excessive Laplace pressure ($p = 2\gamma/r$) of up to hundreds kilobars. Nanoparticles have a highly distorted lattice that has an effect on activation energy of most of the processes they are involved in, by changing their nature and sequence [8]. Nanoscale implies absolutely new effects, properties and processes which are defined by the laws of quantum mechanics, size quantization in small-scale structures, surface-bulk ratio, and other factors [9].

Due to exceptionally small sizes and good reactivity of NP particles, it looks promising to use them in metal melts as crystallization centers to enable structure refinement of cast products. As well known, the grains are finer, the better are mechanical properties of cast products produced from these alloys.

However, the existing methods of powder addition into metal melts cannot be applied to nanopowders due to their unique properties as compared to coarse powders. NP readily forms aggregates, oxidize at relatively low temperatures, and are poorly wetted by a melt, which is most important for crystallization centers. Moreover, nanopowders easily form a dusty blend in the air despite their high density, and even sometimes, this blend can be flammable and even explosive. Therefore, a totally new NP addition method is developed.

After considering a number of options, an efficient method of nanopowder introduction into a liquid metal was proposed [10-16]. This method avoids all of the above problems and consists in pressing of 5-9.5mm diameter rods or wires (Fig. 1) that contain aluminum particles (granules and nanopowder). Nanopowder contents in rods is within 1.5-2.7 mass %, the volume added NP to alloys being no more than 0.05%, and sometimes even thousandths of one percent.

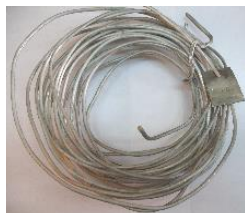

Fig. 1. A standard composition-pressed rod coil (granules of AD aluminum alloy and nanopowder).

The volume of nanopowder in modifying rods depends on the size of particles in the extrudable composition. The larger is the total surface of aluminum particles, the higher is the contents of NP particles in the rod and the less its consumption rate. According to the calculations, the total grit area in the same volume (e.g., in $1cm^3$) at a 0.21mm average diameter is $149cm^2$, which is 21 times as high as that of the granules (average diameter – 2.25mm).

**EXPERIMENTAL.**

The rods of 5mm diameter were pressed from A995 aluminum granules and from A5 aluminum grit and various types of nanopowders. The NP contents in grit rods reached 4-5 mass %, which is 2.0-2.5 times as high as in the granule rods.

Nanopowder volume, which would be sufficient for modification, was determined for AlSi12 alloy. A finished alloy was cast at $720^0C$ into a 400g graphite crucibles pre-heated to $400^0C$, each of them being added with various length modifiers of ⌀5.0mm rod, which was pressed from granules-NP, grit-NP and cracked grain-NP compositions. Alloys were poured into separate cast molds for further testing for mechanical properties. As a result, an optimal NP dose was within 0.004-0.05 mass %.

**RESULTS.**

Table 1 shows the results of modification of D16 distorted aluminum alloy with nanopowders of various chemicals. As seen, the rate of refinement of initial alloy grains by NP addition into the melt reaches 1.7-2.6. Similar results were also attained at semi-solid casting of various sizes round and rectangular ingots of aluminum deformed alloys and cast alloys, cast iron and steel shaped castings, and also when producing aluminum alloy articles by liquid forging [17, 18].

Table 1. Nanopowder effect on average grain size (mm$^2$) in the cross section of D16 die cast, $\varnothing$ 60mm, H = 300mm

| Base metal casting | Granulated metal casting | Granules and Nanopowder-Pressed Rod | | | | | | |
|---|---|---|---|---|---|---|---|---|
| | | $Si_3N_4$ | SiC | $V_{0.75}N_{0.25}$ | $Cr_3C_{1.6}N_{0.4}$ | $B_4C$ | TaN | SiC + $SiO_2$ + Si |
| 0.24 | 0.18 | 0.14 | 0.13 | 0.13 | 0.12 | 0.10 | 0.10 | 0.09 |

Despite a large number of publications describing the ability of refractory NP to refine the structures of metals and alloys, the mechanism of crystallization is not clear understood.

In work [19], an attempt has been made to develop a mathematical model of nanoparticles nucleation effect by assuming that hypothetical $C_mB_m$ NP have a hypothetical spherically symmetric form, with "a model alloy" of a hypothetical $A+B$ composition used as an alloy. An additional assumptions have been also made to simplify the calculations. As a result, the following equation was derived to describe "complete modification" conditions at various doses of the nanopowder:

$$f_s = \frac{pv}{100}\left(\frac{r}{y_0}\right)^3 \qquad (1),$$

where:

$p$ – is powder share, mass %, (0.01 – 0.1 mass %);

$v$ – is alloy – powder density ratio (not indicated);

$r$ – is particle curvature;

$y_o$ – is particle initial size.

The authors believe that complete modification occurs at $f_s > 1$, i.e., the entire crystallized metal structure has a globular shape. At $f_s < 1$ the modification is incomplete, i.e., it results in a mixed structure, the $1 - f_s$ portion of the alloy being crystallized as a conventional shape.

On the other hand, our study states that the efficacy of refinement effect of all nanopowders used (see Table 1) [i.e., AlN; $Al_2O_3$; $B_4C$; BN; $Cr_3C_{1.6}N_{0.4}$; $HfB_2$; HfN; $LaB_6$; SiC; $Si_3N_4$; TaN; TiCN; TiCNO; TiN; $TiO_2$; VCN; $ZrB_2$, and their mixtures (AlN + TiN); (BN + $B_4C$), (SiC + $B_4C$), (SiC + $SiO_2$)] is basically the same. All of them have a similar modifying effect despite great differences in structure and the chemical composition of nanopowders, type and class of lattice, symmetry elements, space group, type of structure, lattice constant, density, melting temperature and other factors.

Nanopowder particles have an exceptional sedimentation ability because of their small sizes (up to 100nm) and a large specific surface area due to the A. Einstein definition in 1905 [20]. According to Einstein, Brownian movement energy would be sufficient for up to 1μm particles to stay at a constant motion without being deposited by gravity. NPs therefore have a double modifying effect. First, they serve as crystallization centers; and second, due to their high volume and a long-term suspension state, they protect the relevant atoms (clusters, blocks) from being diffused with emerging and growing crystals, which eventually promotes the formation of a fine-crystalline structure of cast products. Moreover, as described in the work and mentioned in [21], NPs are capable to strength intermetallic compounds formed in the melts.

Lattice parameters of all refractory compounds differ from those of aluminum and aluminum alloys by more than 10-15% (a face-centered cube with 4.04nm lattice parameter). It can be assumed that nucleation of crystallization centers starts from aluminum monolayer, as formed by NPs when pressing the composition (aluminum particles + NP particles).

CONCLUSIONS.

In the presented work the crystallization mechanism during metal solidification has been proposed. It was found that the chemical composition of the NPs has no influence to the formation of the metallic structure. The modification effect was attribute to particle size added to the melt.

REFERNCES.